\documentclass[aip,amsmath,amssymb,reprint]{revtex4}
\usepackage[dvips]{graphicx}
\usepackage{dcolumn}
\usepackage{color}
\usepackage{ulem}
\usepackage{mathptmx, helvet, courier}
\usepackage{bm,amsmath,amssymb}
\usepackage{threeparttable}
\usepackage{multirow}
\usepackage{array}
\usepackage{etoolbox}

\begin{document}

\title{Self-assembly of model proteins into virus capsids}

\author{Karol Wo{\l}ek}
\affiliation{Institute of Physics, Polish Academy of Sciences,
 Al. Lotnik{\'o}w 32/46, 02-668 Warsaw, \\ Poland}

\author{Marek Cieplak}
\email{mc@ifpan.edu.pl}
\affiliation{Institute of Physics, Polish Academy of Sciences,
 Al. Lotnik{\'o}w 32/46, 02-668 Warsaw, \\ Poland}

\date{\today}

\begin{abstract}
\noindent
We consider self-assembly of proteins into a virus capsid
by the methods of molecular dynamics.
The capsid corresponds either to SPMV or CCMV and is studied
with and without the RNA molecule inside. The proteins are 
flexible and described 
by the structure-based coarse-grained model augmented by 
electrostatic interactions. 
Previous studies of the capsid self-assembly
involved solid objects of a supramolecular scale, e.g. corresponding
to capsomeres, with engineered couplings and stochastic movements.
In our approach,
a single capsid is dissociated by an application of a high temperature
for a variable period and then the system is cooled down to allow for
self-assembly. The restoration of the capsid proceeds to various
extent, depending on the nature of the dissociated state, but is 
rarely complete because some proteins depart too far unless
the process takes place in a confined space.
\end{abstract}

\maketitle

\section{Introduction}

Protein aggregation is ubiquitous and results in different
outcomes depending on the nature of the interactions between the
proteins. Through  cyclization and dimerization of the same protein, 
as well through combination with another protein, aggregation leads to
a finite number of predicted topologies  of protein
complexes in quarternary structure space \cite{Teichmann}.
14 of these topologies are observed in the Protein Data Bank~\cite{pdb}.
Aggregation may generate amorphous clusters during
{\it in vitro} misfolding if there is no protection provided by chaperons
\cite{gething}, but it may also produce quasispherical hollow shells as in the
case of apoferritin \cite{ferritin}. In addition, it may lead to  formation of 
fibrous structures, such as amyloid fibers \cite{Shea,Derreumaux,Knowles,Ranganathan}
or polymers made of sickle cell hemoglobins \cite{Hofrichter}. Finally, a
spontaneous protein aggregation around a nucleic acid \cite{Adolph} 
creates compact virus capsids. The key mechanism for co-assembly of capsid
proteins and RNA is provided by non-specific electrostatic interactions
between RNA phosphate groups and positively charged residues, often located
in flexible tails known as arginine rich motifs \cite{Schneemann}.
There is evidence that there are specific packaging sites on RNA
that additionally affect the process \cite{Dykeman}. It should be
noted, however, that virus capsids can form ({\it in vitro})
without any nucleic acid as a
result of manipulation of the pH of the solvent \cite{Fraenkel}.\\

All of these aggregation processes are difficult to study through
molecular dynamics especially because the entropy significantly
disrupts the proper binding of the assembling units.
Here, we propose an approach in which the fully assembled system
is dissociated in a controlled manner by heating and then 
cooled back to the room temperature in an attempt to restore
the original structure.
Clearly, too much heating will disperse the components too
much for them to reassembly within acceptable computational times. 
Thus there is a threshold bellow which the self-assembly
still takes place, perhaps not fully, and, in this regime, one
can study the reassembly pathways in a meaningful manner. In this
paper, we explore this problem in the context of virus capsids.\\

Most of the quasispherical virus capsids are of the icosahedral
symmetry. The proteins (called subunits in this context) in
such capsids become arranged in special motifs. Here, we consider
self-assembly of icosahedral virus capsids from proteins
that are described by a coarse-grained
structure-based model. This kind of the protein-based
representation of the capsids has been
used previously to study nanoindentation
of the capsids that have been already formed \cite{virus,virus1}.
We focus on two viruses: SPMV 
(satellite panicum mosaic virus) \cite{Scholthof}
and CCMV (cowpea chlorotic mottle virus) \cite{Markham,Trylska}.
CCMV is one of the best studied viruses \cite{virology}.
It contains RNA and 180 identical protein subunits.
The subunits are arranged into 12 pentamers and 20 hexamers,
known collectively as capsomers. This virus corresponds to
the triangulation number, T, of 3 \cite{Speir,Johnson}. 
It is made of 34 200 residues out of which 5 580 belong to 
disordered tails. SPMV is one of the smallest
capsids and its symmetry corresponds to T=1  \cite{McPherson}.
It is made of 9 420 residues grouped into 60 subunits.
960 of these residues are in the tails.\\


The kinetic pathways
of the capsid self-assembly are diversified and the
role of the nucleic acids in the process appears to depend on the
system. An equilibrium Landau-type theory \cite{Rudnick}
suggests that the icosahedral state is in close competition
with states that have tetrahedral and octahedral symmetries which
may confound assembly.
There is experimental evidence that in the case of CCMV the
proteins tend to first form dimers and the capsomers arise by
aggregation of the dimers \cite{Zlotnick}. However, HK97 seems
to form heksamers and pentamers in one step \cite{Hendrix}.
Other experimental insights into the assembly process are scarce
which calls for a thorough analysis of the process
through modelling.\\

Existing theoretical studies of the problem involve
coarse-grained models that use stiff objects imitating
supramolecular objects such as capsomers that may 
correspond to  hundreds of amino acid residues \cite{Endres}.
In particular, Wales \cite{Wales} and Johnston {\it et. al} \cite{Doye}
represent capsomers by rigid pentagonal pyramids so that
the T=1 capsids are made of 10 pyramids.
Interactions between
the apex points are repulsive and those between the base points
are described by the Morse potential. The authors demonstrate
existence of kinetic traps and a hysteretic behavior. A more
detailed model has been considered by Elrad and Hagan \cite{Elrad1,Elrad}.
It involves truncated-pyramidal shapes constructed out of
rigid polymers (see also refs. \cite{Rapaport1,Rapaport,Nguyen})
that minimize their interaction energy in a perfect T=1
icosahedron. Each such object is meant to represent a trimer
of proteins so the well formed capsid consists of 20 stacked objects.
This model has allowed for identification of several characteristic modes
of self-assembly in the presence of a polymer that depend on the
strength of the object-object interactions relative to the
interactions with the polymer (see also refs. \cite{ZlotWang,Gelbart}).
In still another approach \cite{Schwarz}, the capsomers are represented
by hard spheres to demonstrate that the dynamic influx of the capsomers
in a cellular environment facilitates self-assembly.\\

It is natural to adopt a protein-based description of virus capsids
when considering all-atom models \cite{Freddolino}. However, the 
large number of the degrees of freedom involved has allowed only
for short-time assessment of the fluctuational dynamics around
the native, fully assembled conformation. Thus self-assembly, 
necessarily involving conformational changes of the proteins,
needs to be described in terms of a {\it flexible} coarse-grained model.
Here, we consider self-assembly of such proteins. They
evolve according to the Newton's equations of motion whereas
the rigid supramolecular objects, considered in the previous
theoretical approaches,  usually undergo
purely stochastic displacements (though a Newtonian approach has been
proposed in ref. \cite{Chandler}). 
Each effective atom in our model  represents an amino acid residue and
the contact interactions between them are of the Lennard-Jones kind.
The presence of the interactions is determined through
atomic-level considerations  whereas in the models with the
supramolecular solid objects, the intra-object interactions
are engineered.\\

In our previous studies of nanoindentation within the same model
\cite{virus,virus1} (see also ref. \cite{Dima}),
we have observed the crucial role
of the inter-protein contacts in the capsid collapse,
demonstrated existence of large differences in the deformation field
compared to the continuum shell model \cite{Gibbons07}, and related the
Young modulus to the average contact number that a residue is 
a part of. The more detailed description of the model necessitates
making simplifications in the physical setup. Instead of having
a system of diffusing stiff capsomers that would allow for
formation of tens of capsids, we just consider a single capsid.
We separate the capsid into its proteins by an application of heating
and then study the kinetics of self-assembly by restoring the room
temperature. We study empty capsids and capsids with the polymeric RNA.\\

We find that the flexible and structure-based coarse-grained
model of the proteins leads to self-assembly of the capsid in a way
that does not necessarily proceeds through the formation
of capsomers that would then combine into the capsid. It is
the individual proteins that appear to be the agents of the process.
The presence of the RNA molecule is observed to destabilize
the capsid in a slight way, but not to affect aggregation in
a significant way.
The outcome of self-assembly is controlled by the unfolding
temperature, the length of time during which unfolding
is induced, and the waiting time as measured from the
instant at which the room temperature is restored.
Substantial thermal unfolding leads to only a partial reconstruction
of the capsid in the cooling stage. We expect, however,
that applying our procedure to many capsids, instead of just one,
especially under the conditions of confinement,
would improve the quality of self-assembly because a protein
that separates from its original capsid through diffusion
is likely to contribute to construction of
another capsid elsewhere.

\section{Methodology}

The model of the empty capsid is described in refs. \cite{virus,virus1}.
It is a generalization of the approach adopted in studies of
individual proteins as outlined in refs. \cite{JPCM,plos,models}.
The proteins are represented by effective atoms located at the
$\alpha$-C atoms of each residue and the solvent in implicit.
The time evolution is defined
in terms of molecular dynamics with the Langevin noise
representing the influence of the solvent. The noise corresponds
to temperature $T$. The interactions between the effective atoms
divide into those corresponding to the native contacts and to
the non-native contacts. The latter are softly repulsive
and they operate at distances smaller than 4 {\AA}.\\

The native contacts are described by the Lennard-Jones potentials
of depth $\epsilon$ and with the length parameter $\sigma$
determined from the native distance between the corresponding
$\alpha$-C atoms. 
Non-uniform values of $\epsilon$ within proteins have been
demonstrated not to improve the model in any significant manner when
confronted with the experimental data on stretching \cite{models}. 
The value of $\epsilon$ has been calibrated to be equal to about
110 pN~{\AA} \cite{plos} which is close to 1.5 kcal/mol obtained by
matching all-atom energies to the coarse-grained expressions \cite{Poma}.
The room temperature, $T_r$, corresponds to $k_BT$ of 0.3 -- 0.35 $\epsilon$
and in the simulations, we take $T_r=0.3 \; \epsilon /k_B$
($k_B$ is the Boltzmann constant). Temperatures around $T_r$
correspond to the shortest folding times in
the model with the chiral backbone stiffness~\cite{wolek2}
that is used here. After we disassembly the virus by an application
of a high temperature, $T_h$, we attempt to recombine it
by restoring the temperature back to $T_r$.
In our model, we take $T_h$  to be usually of order $1\; \epsilon /k_B$.
Such values reduce the computational time scales, but it should be 
noted that the experimental melting temperatures of virus capsids
are much lower. They are typically in the range 60--80$^o$ C \cite{Raya}.\\

In order to identify the native contacts,
we read in the structure file for the full capsid that is stored
in the VIPERdb data base \cite{vdb}. 
The contact map does not include the disordered
tail segments of the proteins.
We use the overlap criterion (for a fuller discussion 
of possible  contact maps see ref. \cite{Wolek}) to determine the
existence of a native contact between two residues. The contact
is considered to be present
if there is at least one pair of heavy atoms whose enlarged van
der Waals spheres overlap. The radii of the spheres are taken from
ref. \cite{Tsai} and then they are multiplied 
by 1.24 to account for attraction \cite{Settani}. This factor
corresponds to the inflection point in the Lennard-Jones potential.
This leads to 71 520  native contacts in CCMV and 25 980 in SPMV.
They split into intra- and inter-protein contacts. There are
19740 intra- and 6240 inter-protein contacts in SPMV.
In CCMV, the corresponding numbers are 54600 and 16920.
In both cases, the number of the intra-protein contacts is about
three times larger than the number of contacts between the proteins.
Any conformation of the system of aggregating proteins can be
characterized by the fraction of the established contacts
relative to the native numbers of the contacts. We introduce
parameters $Q$, $Q_p$, and $Q_{pp}$ which are the fractions
of all of the contacts established, contacts established within
proteins, and contacts between proteins respectively. 
A contact is considered established if the corresponding distance
between the $\alpha$-C atoms does not exceed $1.5\;\sigma _{ij}$.
This distance exceeds the inflection point in the potential
by $\frac{1}{4} \sigma_{ij}$,
but its precise choice has no dynamical
consequences as it is used merely for descriptive purposes.\\


The simulations are performed in a free space, i.e. without any bounding walls.
The implicit solvent used quenches any ballistic motion of the
atoms and the characteristic time scale, $\tau$, is of order 1 ns.
This is the time needed for the effective atom
to cover a distance of 5 {\AA} through diffusion \cite{Szymczak}.\\

The model outlined above does not include the RNA or the disordered
N-proximal segments in the proteins which contains an ARG-rich RNA
binding motif. Deleting the segments does not inhibit packaging of
the RNA but induces structural changes in the capsid \cite{Rao}.
The structure file 1CWP for the
CCMV protein does not contain entries for the first 41 (chain A)
or 26 (chains B and C) out of 190 residues,
which shows dependence on the location in the capsomer. 
These are the tail segments mentioned in the Introduction.
The structure file
1STM for SPMV does not specify coordinates for the first 16 out of 157
residues. In the improved model, we describe the disordered segments
as chains of residues that are endowed with the excluded volume
but are not capable of forming attractive contacts.\\

All non-neutral residues come with with the electric charges, $q_i$.
In units of $e$, these are --1 for ASP and GLU, +1 for ARG and LYS,
and +0.5 for HIS (to account for the different coexisting 
protonation states of this residue). In addition,
each N-terminus is ascribed the charge of +1 and C-terminus of --1.
There are also charges of --1 on the phosphorus atoms of
each of the bases of the RNA and the RNA itself
is represented as a chain of harmonically connected 
beads separated by a distance of 5.8 $\AA$.
The distance associated with soft repulsion between the beads is 
taken after Voss and Gerstein \cite{Voss} to be 8 {\AA},
{\it i. e.} twice as large as the one associated with the
amino acid residues. The distance of soft repulsion between 
the RNA and amino acid residue beads is 6 {\AA}.\\


The electrostatic interactions are described by the Debye-Hueckel
potential:
\begin{equation}
V^{el}_{ij}(r)=q_iq_j\frac{exp(-r/\nu)}{4{\pi}Dr},
\end{equation}
where $r$ is the distance between the charges, $\nu$=10 {\AA} is the
screening length, and $D$=80 is the dielectric constant of water.
The electrostatic terms do not apply to the pairs of residues
which are already connected by the native contacts because such
connections are generally expected to incorporate electrostatics.
They act primarily within the RNA and between the RNA and the 
charged amino acid residues, especially in the dangling
ends, to which no native contacts can be assigned.\\

The genome content of the CCMV virus has been determined through
mass spectroscopy \cite{Heck}. 
There is a number of different RNA molecules that can be present
in any CCMV capsid. The most common of them are denoted
as RNA1, RNA2, RNA3 and RNA4. Their lengths correspond to
3171, 2774, 2173, and 824 bases respectively.
In most cases, 3 different fraction of capsids are found:
those containing single RNA1 or RNA2 molecules 
and those encompassing both RNA3 and RNA4 molecules.
However, in theory, there are other possibilities for the length
and they range between 100 to 12 000 with the preferential
packaging of about 3 200 \cite{Comas,Sicard} yielding the
the optimal protein/RNA mass ratio of 6:1, 
which allows encapsulations of all RNA in solution.
There is just one molecule of RNA in SPMV and it is made of
826 bases \cite{Masuta}. We adopt a shorthand notation in which
"with RNA", especially in the figures and tables, denotes a model
that takes both the RNA and the protein tails into account.
Otherwise (or with the annotation "empty"), there are no RNA and no
tails as in the previous study \cite{virus,virus1}.

\section{Results}

\subsection{Dependence of the equilibrium properties on the temperature}

\begin{figure}[h]
\centering
\includegraphics[width=0.5\textwidth]{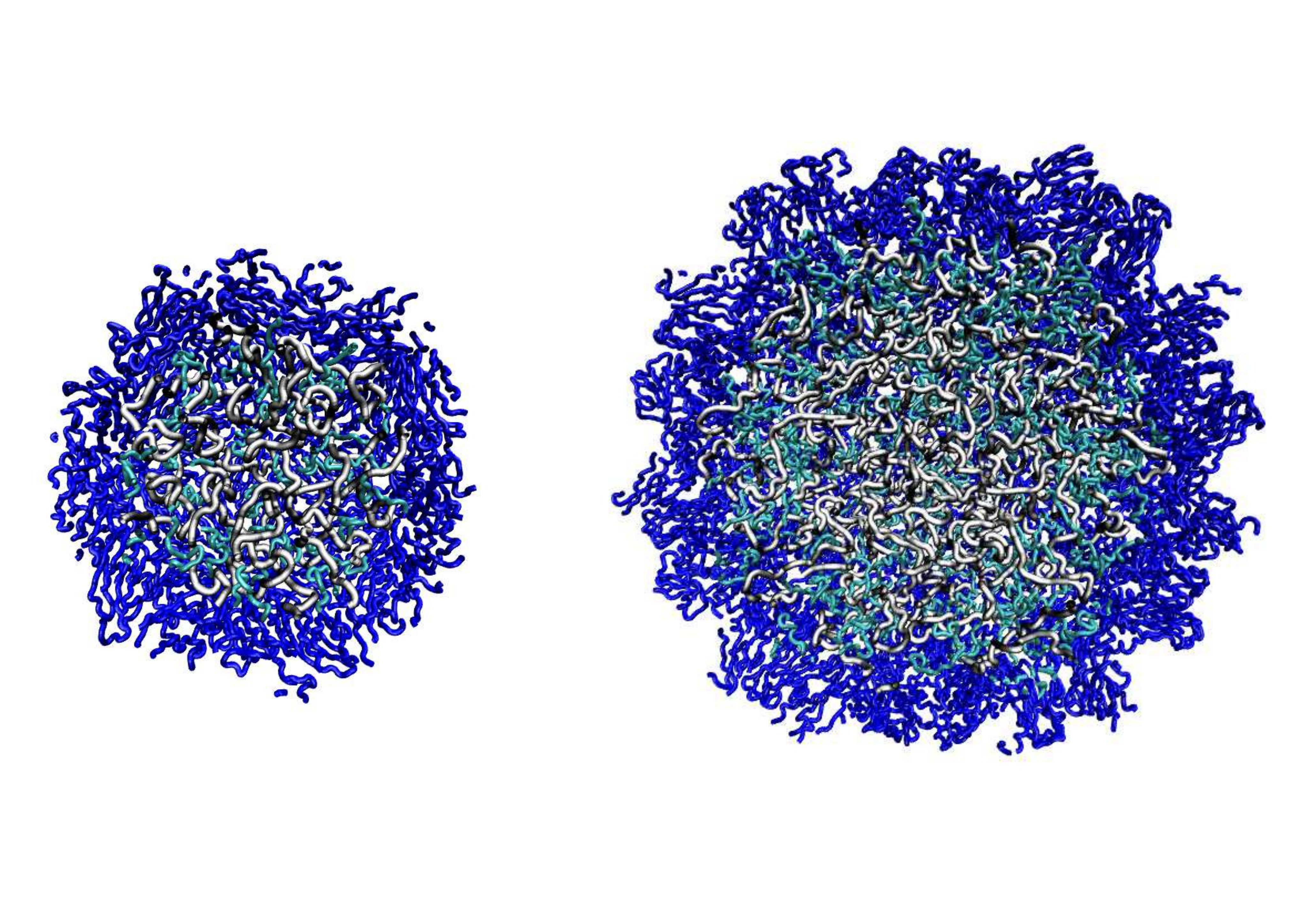}
\caption{Cross-sections of the SPMV (on the left) and CCMV
(on the right) virus capsids in our model. The snapshots are 
shown after equilibration. The dark blue symbols represent the 
structured segments of the proteins whereas the light blue
symbols represent the dangling ends. The RNA molecule is shown in gray.
} \label{photo}
\end{figure}

The initial state of the system with the RNA is derived by starting
with the hollow crystalline structure and adding the missing elements:
the dangling ends and the RNA. These elements are generated as self-avoiding
random walks that also avoid other chains.
When generating such walks, we attempt to select an orientation of each
next bond by choosing random Euler angles up to 10 000 times until
a non-overlapping conformation is found. A failure results in repeating
the construction anew. Such structures
need to be equilibrated at $T_r$. We find that the equilibration lasting 
for 1000 $\tau$ is sufficient. For a meaningful comparison, we
also equilibrate the empty structures in the similar way.
Examples of the derived structures are shown in Fig. \ref{photo}.
They correspond to snapshots obtained at 20000 $\tau$.

\begin{figure}[h]
\centering
\includegraphics[width=0.5\textwidth]{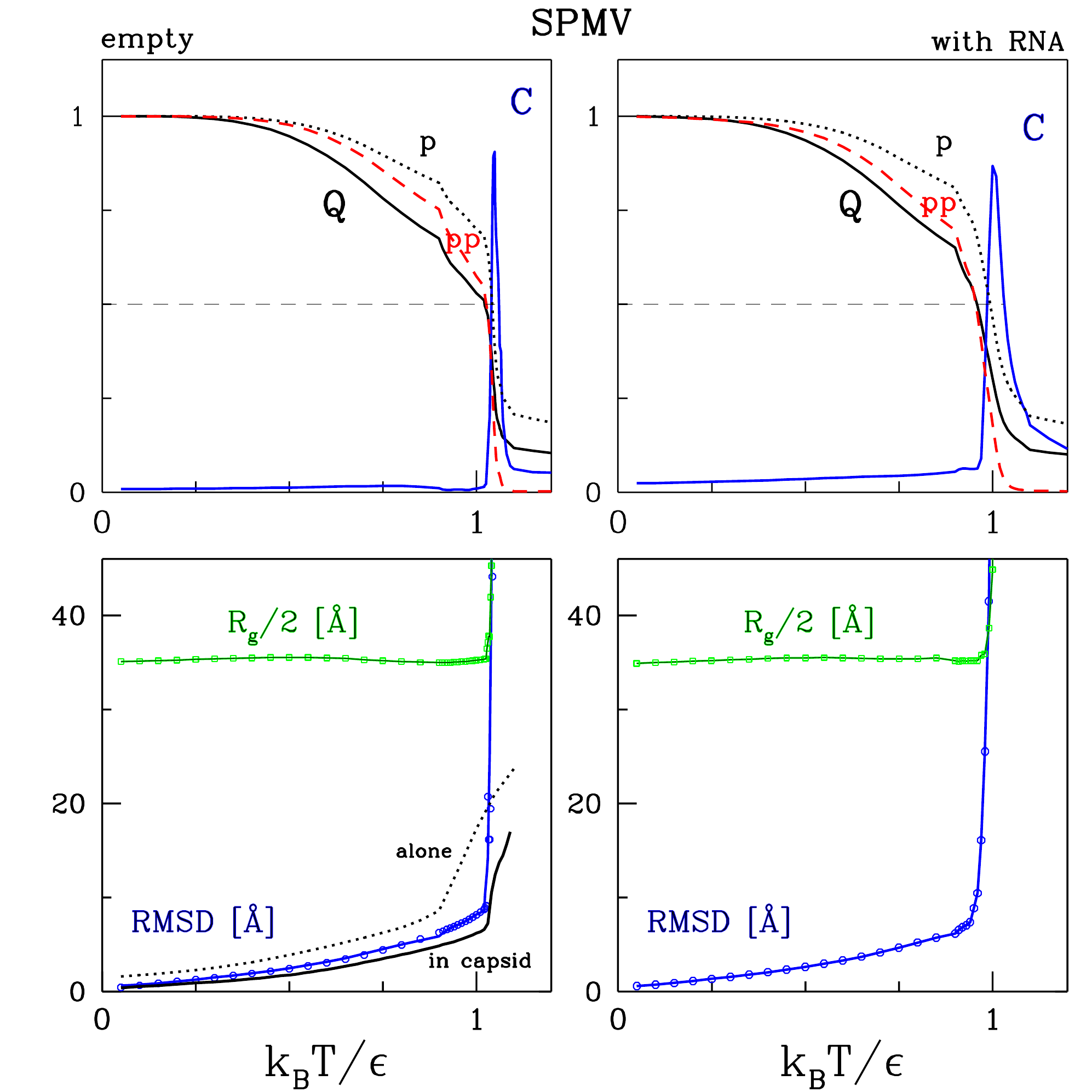}
\caption{The temperature dependence of the equilibrium parameters
describing the SPMV capsid (empty capsid in the left panels and
encompassing the RNA in the right panels). The top panels show the
normalized specific heat (in blue), $Q$ (in black), $Q_{p}$ (in black),
and $Q_{pp}$ (in red). The dashed lines indicate the level of $\frac{1}{2}$.
The bottom panels show $R_g$ and RMSD. The bottom left panel
also shows the RMSD for a single protein when studied alone (the dotted line)
or as a part of the capsid (the solid black line).
} \label{quspmv}
\end{figure}

\begin{figure}[h]
\centering
\includegraphics[width=0.5\textwidth]{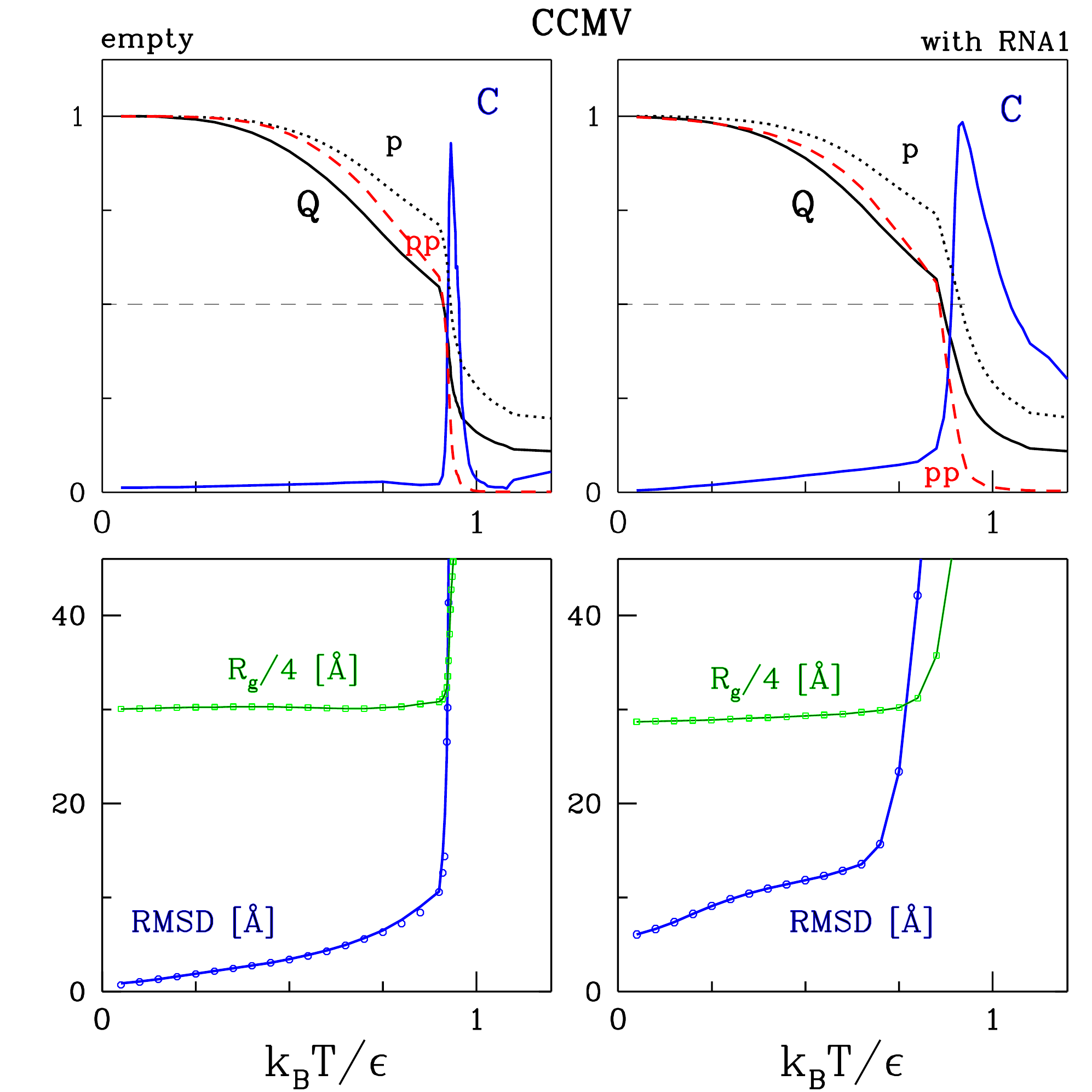}
\caption{Similar to Fig.~\ref{quspmv} but for CCMV, except that
there no results for single proteins are shown.
} \label{quccmv}
\end{figure}

Figs. \ref{quspmv} and \ref{quccmv} show the dependence of the 
equilibrium values of six parameters on $T$ for SPMV and CCMV
respectively, as obtained from 10 trajectories 
of 100000 $\tau$ that start from the conformations generated
through the initial equilibration. The left panels are for the empty
capsids and the right panels are for the capsids with the RNA (in the 
case of CCMV this is the molecule RNA1) and the protein tails.
The first parameter is $C$. This is the specific heat normalized
to its maximal value. The maximum in the specific heat is located
at temperature $T_{max}$, the values of which are listed in Table 
\ref{table1}. Around $T_{max}$ there is a transition between
the quasispherical shape and disordered arrangements.
$T_{max}$ is observed to be lower for CCMV than for SPMV.
The difference is about 10\% both for capsids with the RNA
and without. The presence of the RNA is seen to lead to a 
lowering of $T_{max}$. This happens because the moving  RNA 
molecule keeps striking the capsid shell which contributes
to its destabilization.
The maxima in  $C$ get broader when the RNA is included. The
RNA contributes to fluctuations in the total energy from 
which $C$ is calculated.\\

The other three parameters are $Q$, $Q_p$, and $Q_{pp}$. They 
cross $\frac{1}{2}$ at characteristic temperatures denoted as
$T_Q$, $T_p$, and $T_{pp}$ respectively.
The values of these temperatures are also listed in Table \ref{table1}.
Generally, they are close to $T_{max}$. It should be noted, however,
that the growth in $T$ destabilizes the inter-protein contacts
more than the intra-protein ones. This is reflected in the values
of $T_p$ and $T_{pp}$ and in the plots of $Q_p$ and $Q_{pp}$ in
Figs. \ref{quspmv} and \ref{quccmv}. This is also analogous to 
what happens on squashing the capsid through nanoindentation:
the mechanical collapse of the structure starts by a destruction
of most of the inter-protein contacts.\\

The lower panels in Figs. \ref{quspmv} and \ref{quccmv} also show
$R_g$, the average values of the radius of gyration of the
capsids, and RMSD, the average values of the root mean square
deviations in the positions of the $\alpha$-C atoms relative
to those in the crystalline structure obtained without the
RNA molecules. In the calculation of $R_g$ in the presence of the
RNA, we include the protein tails but not the nucleic acid.
However, in the calculation of the RMSD, the tails do not contribute
as there is no reference structure to compare to.
We observe that both $R_g$ and RMSD grow rapidly around
$T_{max}$. \\

The lower left panel of Fig. \ref{quspmv} also shows the
RMSD for a single protein in two states: in isolation
and as a part of the capsid. We observe that, in the latter case,
the protein is more stable due to the contact interactions
with the neighboring proteins. At $T_r$, The RMSD drops from
2.54$\pm$0.45 to 1.05$\pm$0.10 {\AA} when the isolated 1STM chain is made to
be a part of SPMV. In the case of the 1CWP chain of CCMV, the 
drop is from 3.85$\pm$0.94 to 
1.42$\pm$0.14 {\AA} for chain A and from  7.2$\pm$1.72 to 
1.45$\pm$0.13 {\AA} for chains B and C.\\

\begin{figure}[h]
\centering
\includegraphics[width=0.5\textwidth]{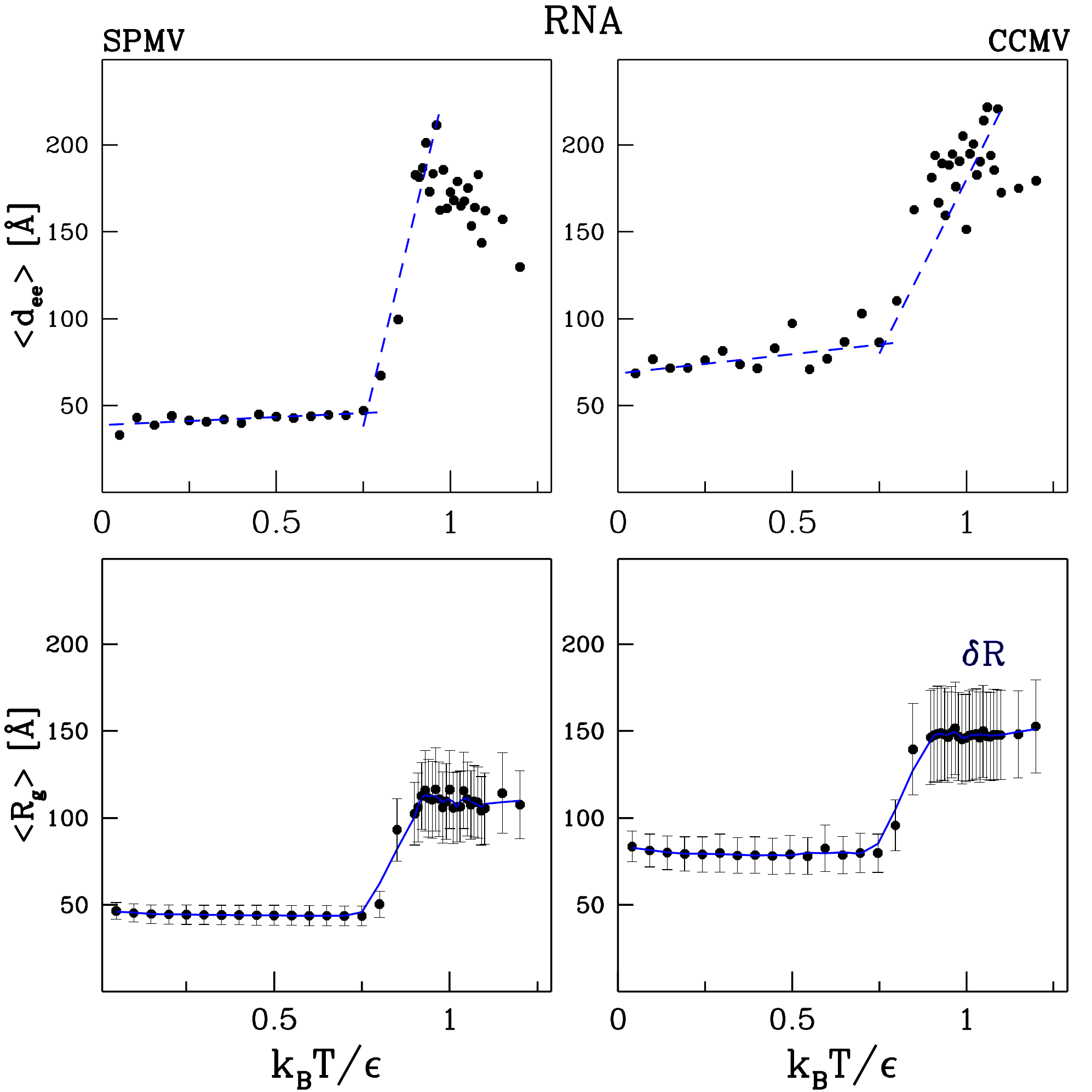}
\caption{Characterization of the RNA molecule in the model SPMV
(the left panels) and CCMV (the right panels) capsids. The top
panels represent the end-to-end distance. The bottom panels
show the average radius of gyration and the vertical bars show the
width of the distribution of the average distance from the
center of mass.
} \label{qurna}
\end{figure}

We now discuss the properties of RNA in a capsid.
Fig. \ref{qurna} shows the $T$-dependence of $R_g$
and the average end-to-end distance, $d_{ee}$, for the
RNA molecule in SPMV and RNA1 molecule in CCMV.
Around $T_{max}$, both quantities are seen to undergo a
rapid rise that is related to the molecule leaving the
dissociating capsid and thus experiencing a significantly
reduced confinement. $R_g$ is observed to switch from a lower
to a higher level on heating. The data points for $R_g$ are very
close to those for $<R>$, which is the average radial distance of
the $\alpha$-C atoms from the (moving) center of mass of the molecule.
The vertical bars in the bottom panels of Fig. \ref{qurna}
show the width, $\delta R$, of the nearly Gaussian distribution of the
distances (the full length of the bars is equal to the width).\\

Table \ref{table2} lists other geometrical parameters
that pertain to the capsid: the average distance from the
center of mass, $<R>$, $R_g$, the width of the radial distribution
of the mass, $\sigma _R$ which serves as a measure
of the thickness of the viral shell, and the average minimal and
maximal distances from the center of mass to the $\alpha$-C atoms.
($\sigma _R$ is analogous to $\delta R$, but the former is for 
the proteins and the latter for the RNA.)
All of these averages are calculated at $T_r$
and compared to the native values whenever the
nucleic acid is absent (for a more extensive discussion
of the native-state geometry of the capsids see ref.\cite{Chwastyk}).
We observe that $<R>$ is very close to $R_g$. With the RNA, 
$R_g$ is smaller than without because of the electrostatic attraction
between the more or less centrally located nucleic acid and the proteins.
In the case of CCMV the reduction in $R_g$ is by
4\%. However, the thickness with the RNA
is larger than without,  because of the tails that tend to point
away from the structured parts of the proteins.
The tails are also responsible for the substantial lowering the 
the values of $R_{min}$. We observe that
the electrostatic attraction between the RNA and the proteins
affects primarily the dangling ends:
when the dangling ends are removed, the values of $<R>$ and $R_g$ are 
found to be nearly the same as in the systems without the RNA.

\subsection{Dissociation of the capsids}

\begin{figure}
\centering
\includegraphics[width=0.75\textwidth]{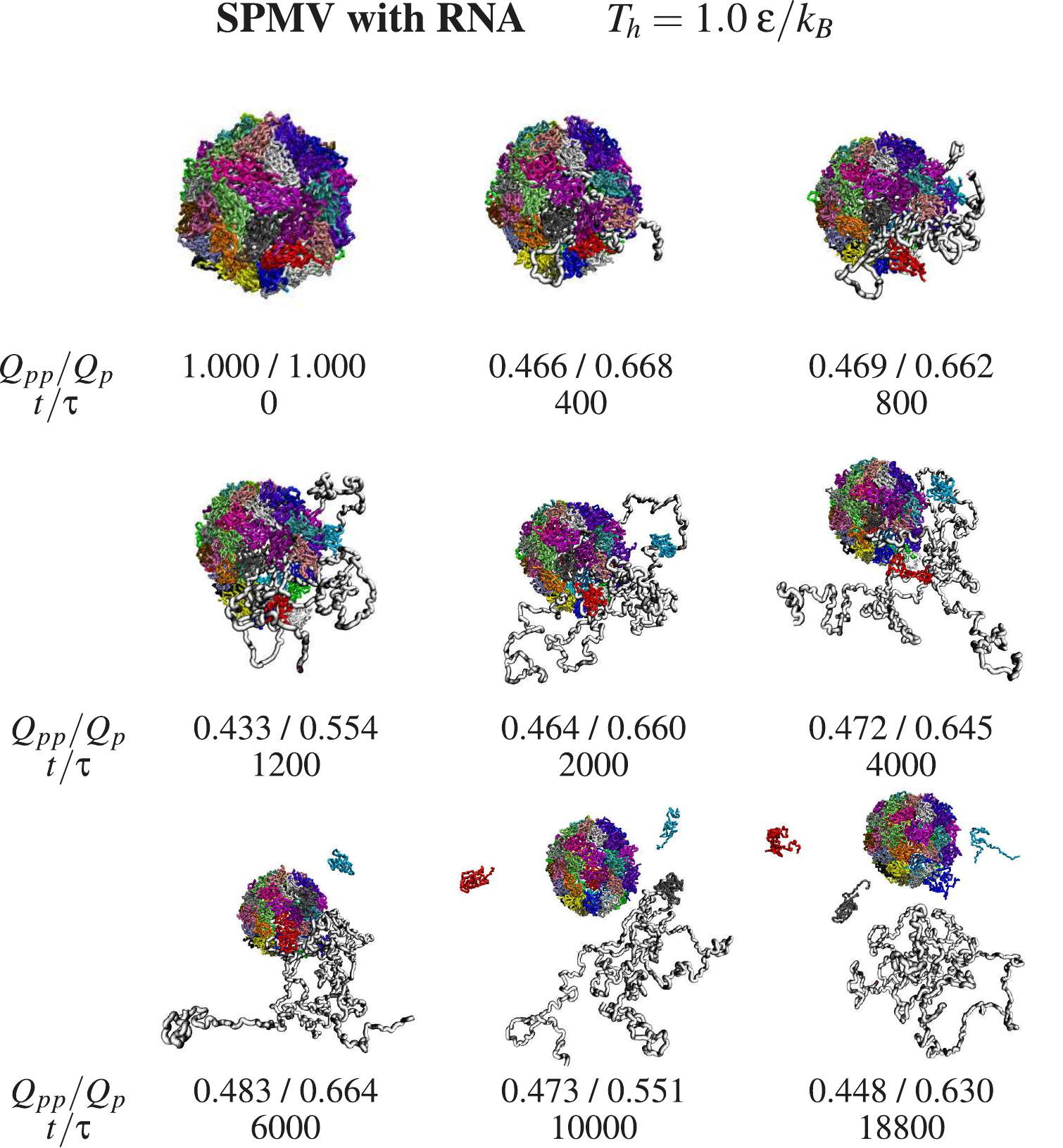}
\caption{Snapshots form one trajectory of dissociation of the model SPMV
with RNA at $T_h$ written at the top. Under each snapshot, there is 
information about the corresponding values of $Q_{pp}$, $Q_p$, and the
time of heating. The colors used to show proteins are arbitrary. The
RNA is shown in gray.
}
\label{heatrnaspmv}
\end{figure}

One may obtain fast dissociation by selecting $T_h$ to be
in the vicinity of $T_{max}$. 
Such temperatures are unrealistically high, but they serve the
numerical purpose and can also be thought of as representing
potent chemical denaturants. Figures \ref{heatrnaspmv} and
\ref{heatrnaccmv} show examples of the dissociation process
for SPMV at $T_h=1.0\; \epsilon /k_B$ and CCMV at
$T_h=0.9\; \epsilon /k_B$, both with the RNA molecule, respectively.
The subsequent conformations are characterized by the values 
of $Q_{pp}$ and $Q_p$. In the snapshots shown for SPMV,
$Q_{pp}$ decreases (not strictly monotonically) from 1 to
0.448 in the time span of 18800 $\tau$.
In the case of CCMV, $Q_{pp}$ decreases to 0.006 in a
comparable time span of 19600 $\tau$. Despite the increasing
number of the ruptured links between the proteins, the proteins
themselves are pretty well connected by the internal 
contacts. In the final stage shown, $Q_p$ is 0.630 for SPMV
and 0.481 for CCMV.
There appears to be an important difference between the
behavior of the RNA molecule in the two systems.
For SPMV, the RNA separates from the capsid proteins
entirely whereas for CCMV, RNA1 continues to be surrounded
by the proteins in all directions. The difference may
have to do with the larger mobility of the four times shorter
RNA in SPMV compared to CCMV, or perhaps also, to the
specific choice of the temperature.\\

\begin{figure}[h]
\centering
\includegraphics[width=0.5\textwidth]{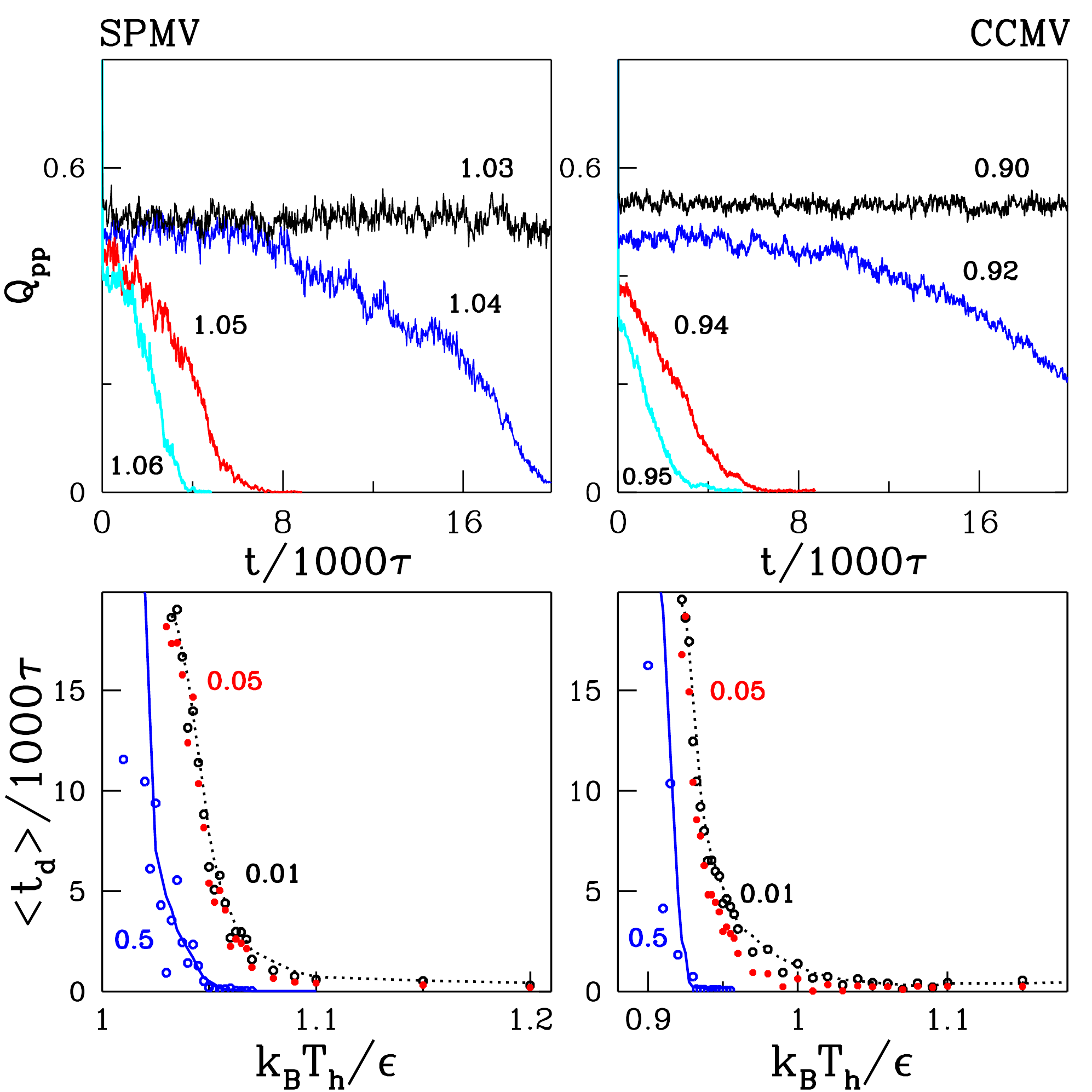}
\caption{The top panels show the time dependence of $Q_{pp}$
during heating, on the left for SPMV and on the right for CCMV.
The numbers indicate the values of $T_h$ in units of $\epsilon /k_B$.
The bottom panels show the average dissociation times for various
indicated levels of what is considered to be a successful dissociation.
} \label{disoc}
\end{figure}

\begin{figure}[h]
\centering
\includegraphics[width=0.5\textwidth]{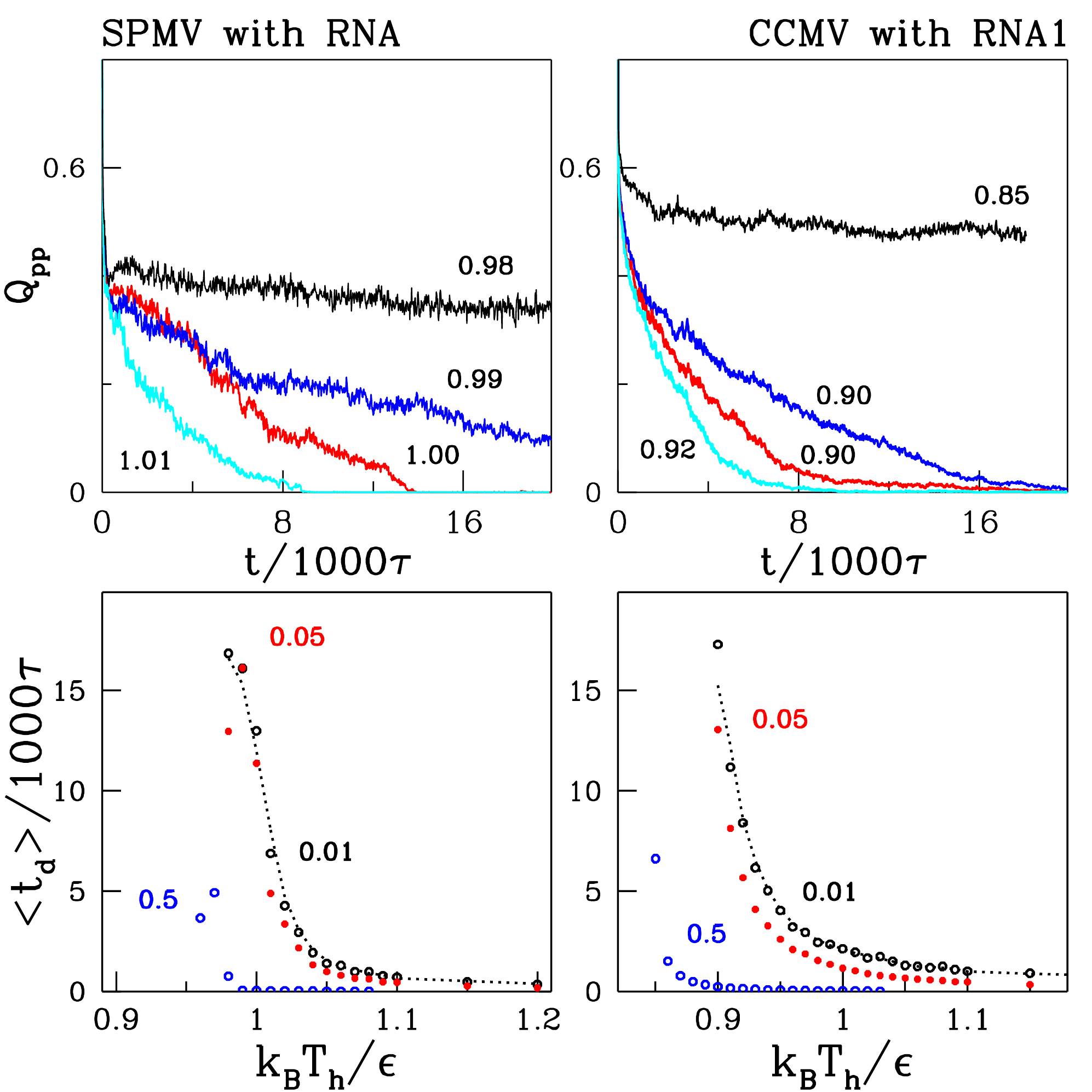}
\caption{Similar to Fig.~\ref{disoc} but for capsids with RNA.
} \label{disocrna}
\end{figure}

The disintegration is a kinetic process and its observed outcome
depends on the value of $T_h$ and the duration of heating. This is
illustrated in the 
top panels of Figs.~\ref{disoc} and \ref{disocrna} 
which show the time ($t$) dependence of $Q_{pp}$ at several 
temperatures in the vicinity of $T_{max}$ for the systems
considered. The second of these figures 
is for the systems with the RNA and the first -- without.
For the $T_h$ selected, the dissociation times, $t_d$, are
of order 1000 -- 10000 $\tau$. \\

Figs.~\ref{disoc} and \ref{disocrna} show the average dissociation
times needed for $Q_{pp}$ to drop
to predefined threshold value, $Q_{th}$, as a function of $T_h$.
The data points are based on 20 trajectories.
We consider $Q_{th}$ to be 0.01, 0.05, and 0.5 as indicated
in the figure. The more stringent the disintegration
criterion is (the lower value of $Q_{th}$), the longer
the corresponding time. Another way to describe the data
in Figs.~\ref{disoc} and \ref{disocrna} is to say that a given dissociation
time is achieved at a higher $T_h$ if $Q_{th}$ is lowered.
By manipulating $T_h$ and the time of heating
we can prepare a capsid corresponding to a given 
value of $Q_{pp}$ and then observe how it aggregates
on restoring the $T$ back to $T_r$.



\subsection{Self-assembly of the capsids}

\begin{figure}
\centering
\includegraphics[width=0.75\textwidth]{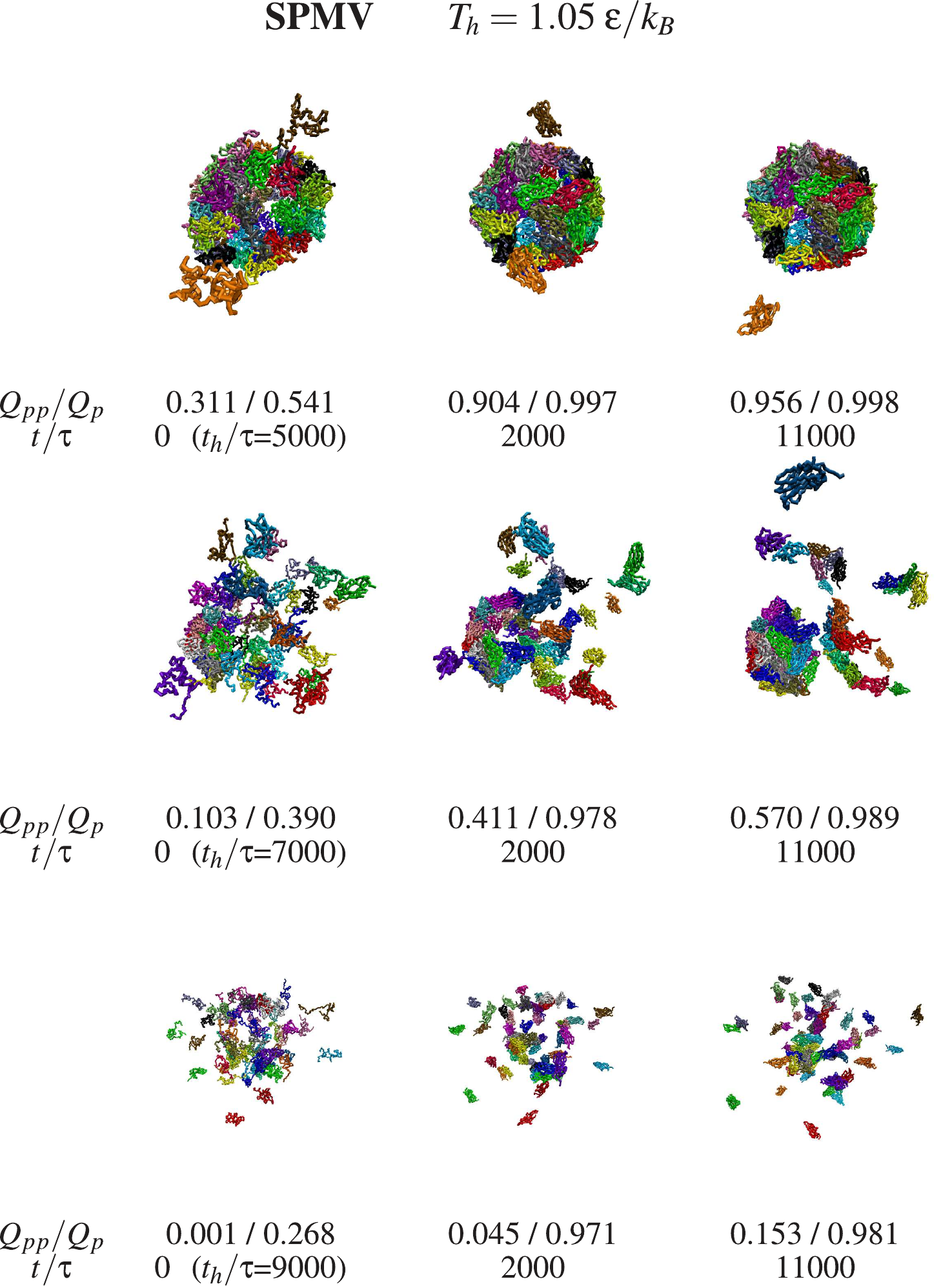}
\caption{Examples of the SPMV capsid assembly after thermal denaturation 
at the temperature indicted at the top. Each horizontal triplet of
panels shows snapshots appearing after evolving from the leftmost
structure. This starting structure has been obtained at $T_h$ applied
for time $t_h$ written underneath in the brackets. The values of 
$Q_{pp}$ and $Q_p$ are indicated. The colors of the proteins are
arbitrary.
}
\label{spmvag}
\end{figure}

\begin{figure}
\centering
\includegraphics[width=0.75\textwidth]{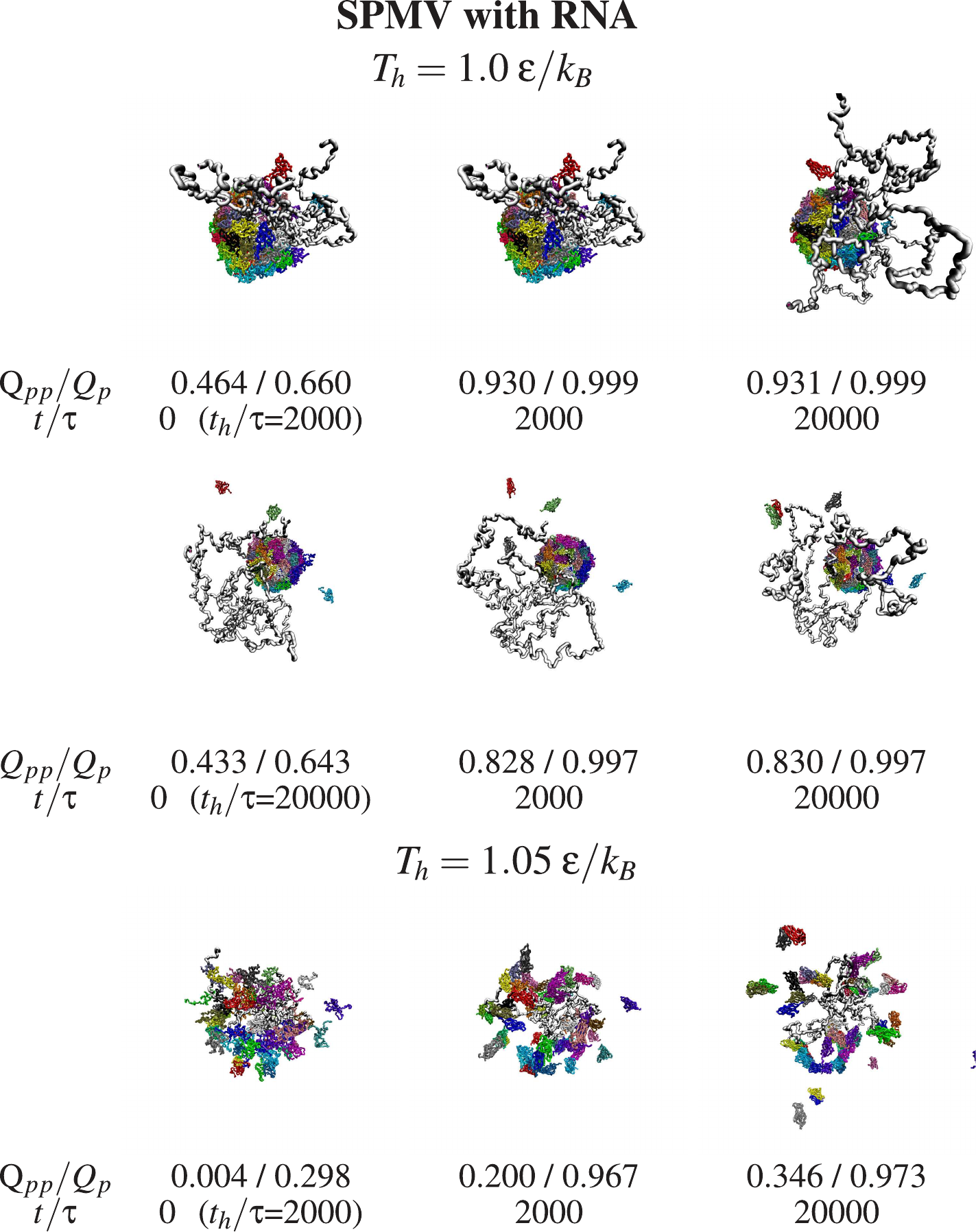}
\caption{Similar to Fig.~\ref{spmvag} but form SPMV with RNA.
The RNA molecule is shown in gray.
}
\label{spmvagrna}
\end{figure}

\begin{figure}
\centering
\includegraphics[width=0.75\textwidth]{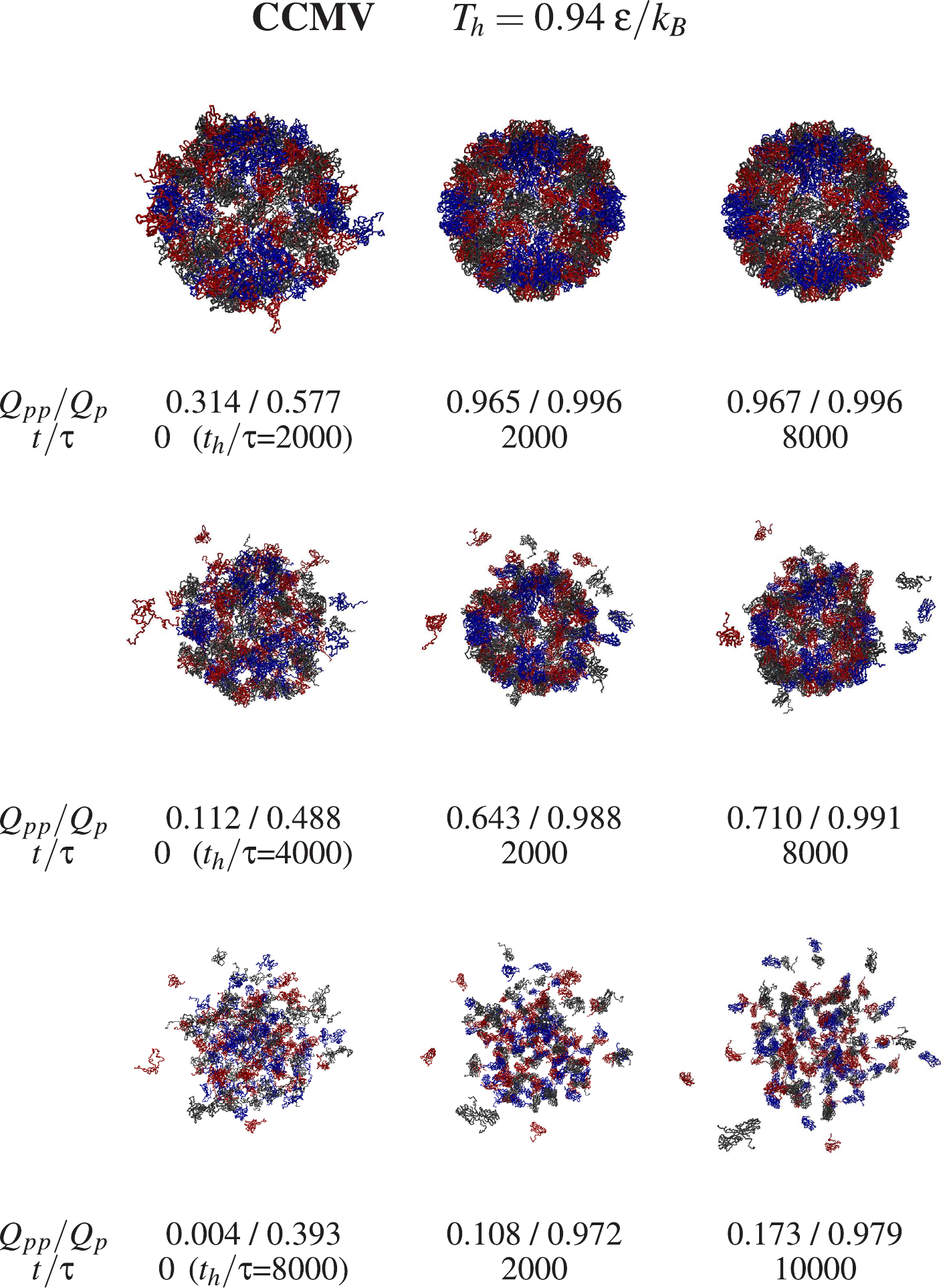}
\caption{Similar to Fig.~\ref{spmvag} but for for CCMV.
Chains A, B, and C are marked in blue, red, and black respectively.
}
\label{ccmvag1}
\end{figure}

\begin{figure}
\centering
\includegraphics[width=0.75\textwidth]{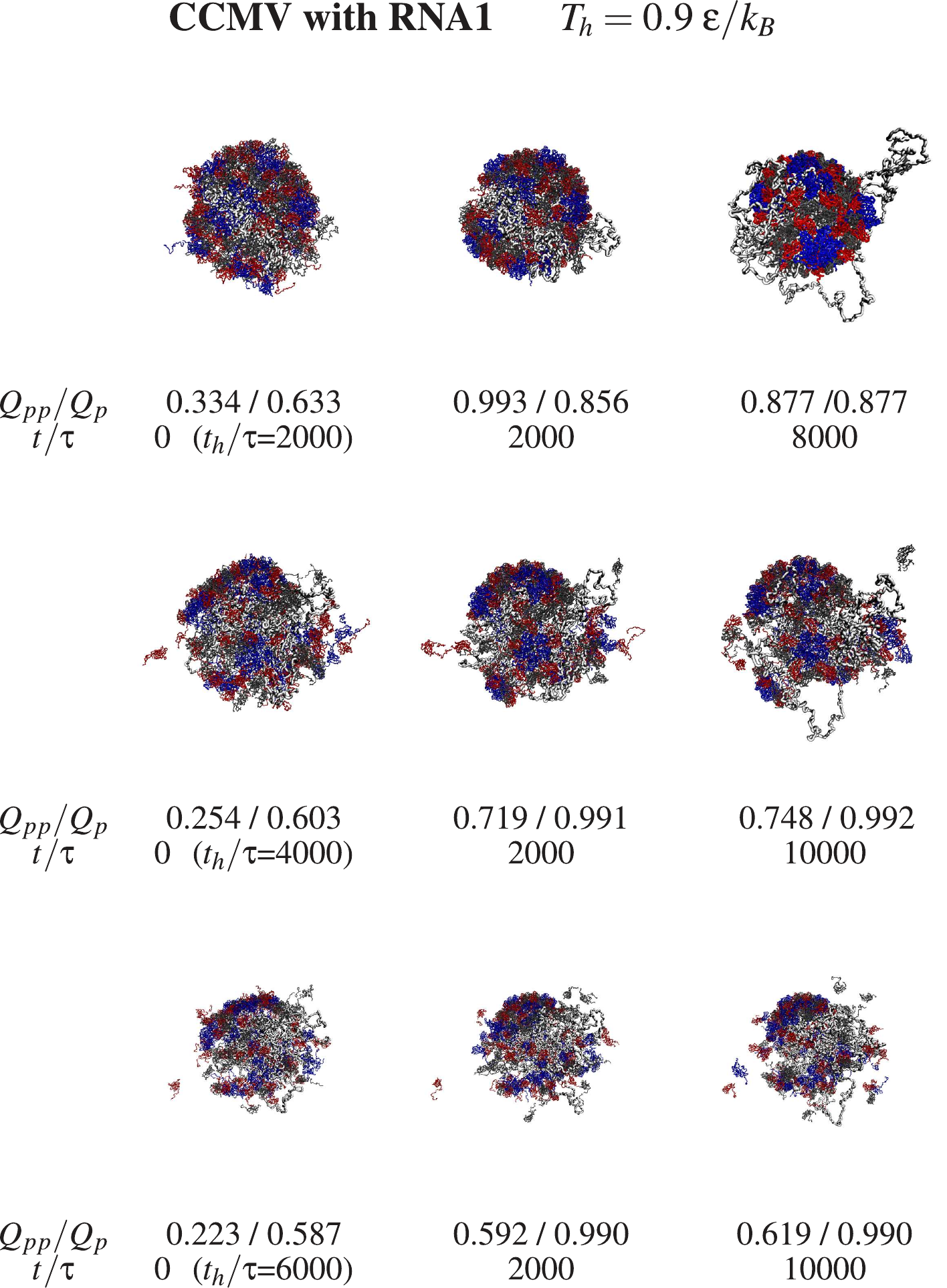}
\caption{Similar to Fig.~\ref{spmvag} but for CCMV with RNA1.
Chains A, B, and C are marked in blue, red, and black respectively.
}
\label{ccmvagrna1}
\end{figure}

We now consider aggregation and discuss what happens with the
dissociated fragments when the temperature is switched back
from $T_h$ to $T_r$. Examples of triplet snapshots from the aggregation
trajectories are shown in Figs.~\ref{spmvag}, \ref{ccmvag1},
\ref{spmvagrna}, and \ref{ccmvagrna1}, where the first two figures
address the systems without the RNA  and the last two -- with the RNA.
In each triplet, the first snapshot defines the state which
is considered to be initial for the studied aggregation process.
This initial state is characterized by the values of $T_h$
(specified at the top of each figure) and the duration of the
dissociation, $t_h$, (specified next to the first snapshot
in each triplet).\\

The snapshots point to a steady growth in the inter-protein
connectivity and to an aggregation which, in the case of SPMV,
leaves the RNA outside of the assembling capsid when the
initial state corresponds to the RNA being separated.
The energy terms in our model do not appear to provide means
of return penetration of the capsid by the RNA. \\

\begin{figure}
\centering
\includegraphics[width=0.75\textwidth]{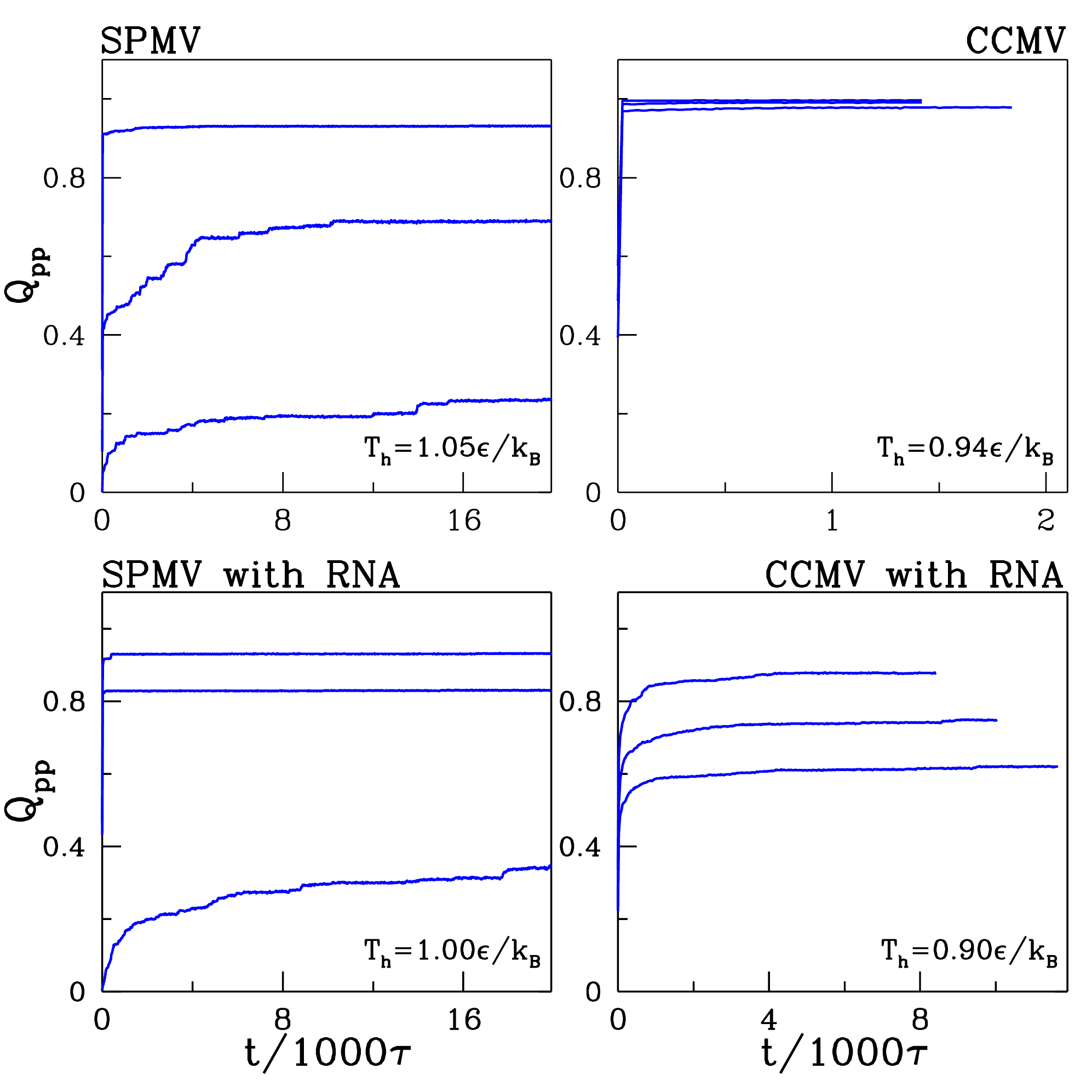}
\caption{The time dependence of $Q_{pp}$ during self-assembly
at $T_r$ for the systems indicted. The initial states were obtained
by heating at $T_h$ with values written in the right bottom corners
of the panels.
}
\label{agreg}
\end{figure}

Fig.~\ref{agreg} shows the $t$-dependence of $Q_{pp}$ in the
trajectories from which the snapshots were captured.
We observe that, at least within our time scales, the self-assembly
is never perfect. $Q_{pp}$ is seen to usually rise rapidly and then
to saturate on a constant value, which may be even as high
as nearly 90\%, but typically is much smaller. The incomplete nature
of the process is primarily due to some proteins departing
too far away from the original location of the capsid that dissociated.
Reconstruction speeds can be defined as the time derivatives of $Q_{pp}$.
Their analysis at short time scales  indicates an approximate
$\frac{1}{t}$ decrease. Based on this, we estimate that achieving the
ultimate saturation level should take several seconds.\\

We do not observe any clear signature of assembly that would
proceed by first forming capsomeres and then joining the capsomeres
together. Heating may disrupt local structural patterns
but they are obviously capsomer related: any group of
proteins may rupture and then come back to the original
state on cooling, if the perturbation is not too large.
Separated proteins may combine into clusters but 
the clusters are not necessarily capsomerial entities.
The proteins do not have identity and appear to act similar 
to condensing gas molecules that can fit to many places in
a growing droplet. \\

It should be noted that
our model is defined primarily by the native contact map.
Thus, when two proteins recombine, it is of a secondary importance
whether they belong to the same or different capsomers,
unless there is a strong imbalance in the number
of the connecting contacts. There could be a difference in statistics,
but we could not capture it. It requires further studies to
figure out whether comparable formation of the intra- and inter-capsomer
dimers is the feature of the structure-based  model or is more
general. It would also be interesting to do a systematic study
for various viruses in this context.

\section{Conclusions}

We have considered self-assembly of flexible proteins coming from a
single capsid that gets dismantled thermally. We have used the
the structure-based coarse-grained model with short range contact
interactions and effectively short range Coulomb interactions.
We demonstrate that this model does lead to self-assembly
but the process is incomplete because of some proteins diffusing
outside of the range of the interactions. The escape of the
proteins could be eliminated by considering the process
under the conditions of confinement.\\

In a situation with many capsids in a solvent, and not just one
considered here, it is possible that a stray protein may dock
properly into some other self-assembling capsid, leading to its
more complete construction. It would be interesting to generalize
our model to a multi-capsid version and to study self-assembly
as a function of the number of the capsids  and
under confinement. It should
be noted, however, that the multiple-capsid problem involves
conceptual issues when considered within the structure-based model.
These issues are not solved yet. For a single globular protein,
the native structure defines a unique contact map (for a given
scheme of selecting the contacts). However, a possibility of
aggregating proteins that belong originally to
various capsids requires defining a contact map which sheds
information about the capsid of origin.\\

A multi-capsid model that needs to be constructed
could also be used to analyze formation of 
capsid lattices on solids, which are of interest in 
biotechnological applications \cite{Yeates,Valbuena}.
Another related direction of a future research within our approach could be
considering virus self-assembly on a fluctuating lipid membrane
\cite{Hagan1} since the membranes can promote association.\\

\begin{figure}
\centering
\includegraphics[width=0.75\textwidth]{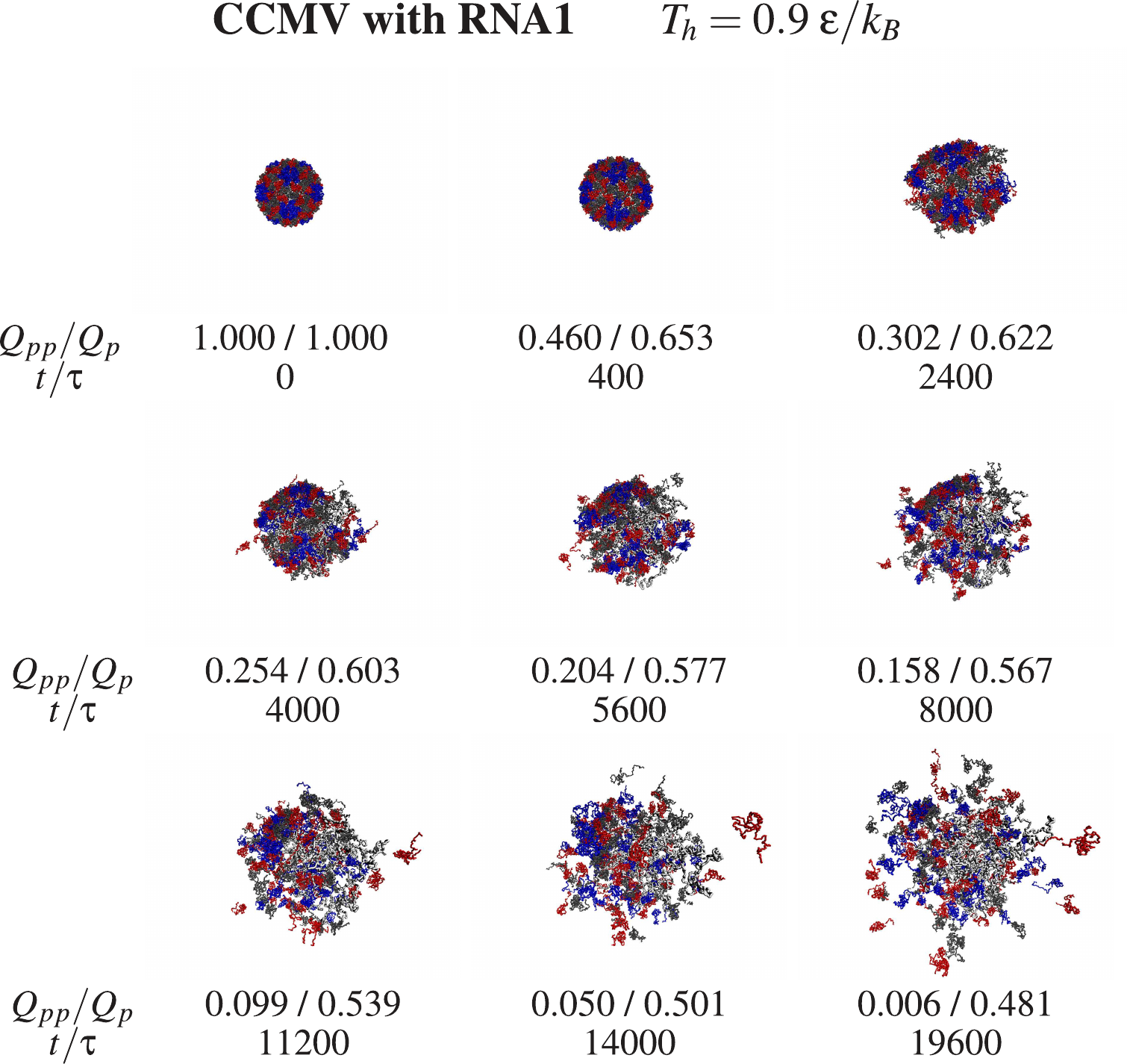}
\caption{Similar to Fig.~\ref{heatrnaspmv} but for CCMV with RNA.
Chains A, B, and C are marked in blue, red, and black respectively.
}
\label{heatrnaccmv}
\end{figure}

We have not observed any clear differences between self-assembly
of SPMV and CCMV except that, during the dissociation
taking place around $T_{max}$,
the RNA molecule finds it easier to leave the SPMV shell than
the CCMV one and then cannot get back inside.
This difference is primarily due to the fact that the RNA molecule
associated with SPMV is much more mobile than RNA1 associated
with CCMV because it is a factor of 4 shorter sequentially.
However, the dissociation patterns are governed also
by the temperature. At temperatures higher than $T_{max}$ the SPMV
capsid fully unravels in a way shown in Fig.~\ref{heatrnaccmv} for
CCMV near $T_{max}$.\\


Our model does not explicitly introduce a possibility of
of hierarchical assembly in which binding characteristics
depend on the stage of the process \cite{Baschek}
(say, forming capsomeres involves different propensity than
that of the full capsids). However, such features may
arise naturally and are worth being explored.\\

{\bf Acknowledgements}
KW was supported by the European Framework Programme VII NMP grant 604530-2
(CellulosomePlus)  which was cofinanced by  by the Polish Ministry of
Science and Higher Education from the resources granted for the
years 2014-2017 in support of international scientific projects.
MC has received funding from
the National Science Centre (NCN), Poland, under grant No.~2014/15/B/ST3/01905.
MC has also benefited from grant No.~2015/19/P/ST3/03541 to Panagiotis
Theodorakis administered by NCN and awarded by
the European Union's Horizon 2020 research 
and innovation programme under the Marie Sk{\l}odowska--Curie grant 
agreement No. 665778.
The computer resources were supported by the PL-GRID infrastructure
and financed by the European Regional Development Fund under the
Operational Programme Innovative Economy NanoFun POIG.02.02.00-00-025/09.

\clearpage

\begin{table}

\begin{tabular}{|c|c|c|c|c|}

\hline
CAPSID &  $\;k_BT_{max}/\epsilon \;$ & $\;k_BT_Q/\epsilon \;$ & $\; k_BT_{p}/\epsilon \;$ & $\; k_BT_{pp}/\epsilon \;$ \\ 
\hline
\hline
{\bf SPMV} & 1.045& 1.021 & 1.044 & 1.029 \\
SPMV with RNA  &1.025 & 0.965 & 0.991 & 0.961 \\
  &    &   &   & \\
{\bf CCMV} & 0.932& 0.904 & 0.929 & 0.908 \\
CCMV with RNA1 & 0.912 & 0.865  & 0.910 & 0.852  \\
\hline
\end{tabular}
\caption{Characteristic temperatures for the systems studied. }
\label{table1}
\end{table}

\begin{table}

\begin{tabular}{|c|c|c|c|c|c|}

\hline
CAPSID & $<R>$ [{\AA}] & $R_g$ [{\AA}] & $\sigma_R$ [{\AA}] & $R_{min}$ [{\AA}] &$R_{max}$ [{\AA}]   \\ 
& native \& at $T=0.3\epsilon/k_B$ & native \& at $T=0.3\epsilon/k_B$& native \& at $T=0.3\epsilon/k_B$ & native \& at $T=0.3\epsilon/k_B$ & native \& at $T=0.3\epsilon/k_B$ \\
\hline
\hline
{\bf SPMV} 	& 69.66 \hspace*{0.3cm} 70.54 & 69.97 \hspace*{0.3cm} 70.84 & 6.64 \hspace*{0.3cm} 6.59 &56.99 \hspace*{0.3cm} 55.76&85.37 \hspace*{0.3cm} 87.79 \\
SPMV with RNA & -- \hspace*{0.9cm} 68.29  & -- \hspace*{0.9cm} 68.96	&-- \hspace*{0.9cm} 9.63 &-- \hspace*{0.9cm} 21.89 &-- \hspace*{0.9cm} 88.10  \\
without dandling ends & -- \hspace*{0.9cm} 70.68  & -- \hspace*{0.9cm} 70.98	&-- \hspace*{0.9cm} 6.52 &-- \hspace*{0.9cm} 56.56 &-- \hspace*{0.9cm} 87.86  \\
& & & & &\\
{\bf CCMV} 	& 119.56 \hspace*{0.3cm} 121.39	& 120.02 \hspace*{0.3cm} 121.84 &10.54 \hspace*{0.3cm} 10.36&95.34 \hspace*{0.3cm} 93.25&142.49	\hspace*{0.3cm} 145.92 \\
CCMV with RNA1 & -- \hspace*{0.9cm} 115.84		& -- \hspace*{0.9cm} 117.03		& -- \hspace*{0.9cm} 16.61  &--	\hspace*{0.9cm} 34.79	&-- \hspace*{0.9cm} 149.76  \\
without dangling ends & -- \hspace*{0.9cm} 121.39		& -- \hspace*{0.9cm} 121.83		& -- \hspace*{0.9cm} 10.36  &--	\hspace*{0.9cm} 96.88	&-- \hspace*{0.9cm} 146.50  \\
\hline
\end{tabular}
\caption{Characteristic geometric properties of the systems studied. 
The equilibrated values are determined from 5 simulation that are 100000$\tau$ long. Without dangling ends means that, they were not consider in parameters calculation but were present in simulation.}
\label{table2}
\end{table}


\clearpage

\end{document}